\documentstyle[11pt,paspconf,psfig]{article}

\begin{document}

\title{Abundances of Heavy Elements and CO Molecules in High Redshift Damped
Lyman-alpha Galaxies}
\author{Limin Lu, Wallace L.W. Sargent}
\affil{Caltech, Astronomy Department, 105-24, Pasadena, CA 91125, USA}
\author{Thomas A. Barlow}
\affil{Caltech, IPAC, 100-22, Pasadena, CA 91125, USA}

\begin{abstract}

Damped Ly$\alpha$ systems seen in spectra of background quasars 
are generally thought to represent high redshift counterparts of
present-day galaxies. We summarize observations of heavy element 
abundances in damped Ly$\alpha$ systems. The results
of a systematic search for CO and \textsc{C~ii}$^*$
absorption in 17 damped Ly$\alpha$ systems are also presented
using observations obtained with the 10m Keck telescopes.
The latter provides a useful constraint on the expected strength of
[\textsc{C~ii}] 158$\mu$m emission from damped Ly$\alpha$ galaxies.
It is hoped that these results will be useful for planning
future radio to millimeter wave observations of high redshift
galaxies using next generation instruments which are now being built.

\end{abstract}


\section{Damped Ly$\alpha$ Systems as the Progenitor of Normal Galaxies}
 
High redshift galaxies identified through their 
absorption line imprints on
spectra of background quasars are remarkable in the sense that
their selection is by their chance 
alignment with the background quasars and hence
do not depend on their emission properties at any specific 
wavelength. Accordingly, any galaxy containing a significant amount of gas
should be picked up without regard to its stellar content (but see
section 6 for possible caveats).
Consequently, these galaxies should be more representative of
the high redshift universe than quasars, AGNs, radio galaxies, 
infrared luminous galaxies, or Lyman break galaxies;
all of which have been detected at high redshifts.

  It is generally agreed that the population of quasar absorption 
systems that are most closely related to the progenitor 
of normal galaxies is the so-called damped Ly$\alpha$ (DLA) systems
(Wolfe 1988). 
The strongest evidence linking DLA systems to the progenitors 
of normal galaxies is that the baryonic mass density contained 
in the neutral gas in DLA systems at redshift
$z\sim 3$ is comparable to the mass density contained in stars in
galaxies today, implying the transformation of gas in DLA systems
to stars in nearby galaxies over the last 10-15 billion years 
(Lanzetta, Wolfe, Turnshek 1995). DLA systems have 
neutral hydrogen column densities
N(\textsc{H~i})$\simeq 10^{20}-10^{22}$ cm$^{-2}$, similar to
those obtained along sightlines through the Galactic disk. DLA
systems are the only class of quasar absorption line systems 
that show 21-cm absorption against radio background sources 
(cf. Briggs 1988). Their low metallicities, low dust contents,
and low molecular contents (see later sections of this article)
are all consistent with properties expected of
galaxies that are in the early stages of evolution. Consequently,
they should be the prime targets for radio and millimeter wavelength
studies of normal galaxies at high redshifts with future
instruments. In this article, we discuss some of the properties of
DLA systems at $z>1.6$ that may be helpful for planning
future radio and millimeter observations.
 
\section{Heavy Elements  in Damped Ly$\alpha$ Systems}
 
\begin{figure}
\centerline{\vbox{ \psfig{figure=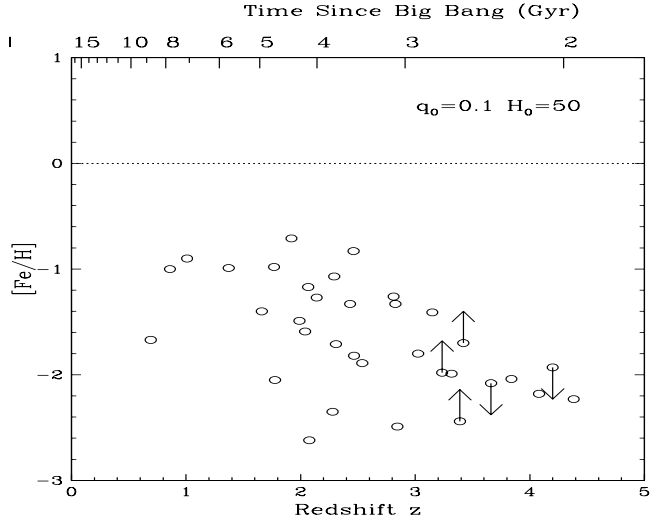,height=3.0in,width=3.5in} }}
\caption[]{Metallicity distribution of damped Ly$\alpha$ systems
as a function of redshift, 
where [Fe/H]$\equiv$log(Fe/H)$_{\rm DLA}-$log(Fe/H)$_{\odot}$ 
is the logarithmic
abundance of Fe in damped Ly$\alpha$ systems relative to the Sun.
The age of the universe starting from Big Bang is given on the top
axis for the assumed cosmological model.}
\end{figure}
 
Abundance determinations for DLA systems began in the late 70's 
(Smith, Jura, \& Margon 1979), but it was only recently that 
abundance estimates for a large number of systems became available
(eg, Lu et al 1996; Pettini et al 1997a,b). In Figure 1
we show the distribution of [Fe/H] vs $z$ for a sample of
DLA galaxies, where the data are taken from Table 16 of Lu
et al (1996) with the addition of
14 new measurements based on unpublished work of our group and
of Wolfe and Prochaska (5 systems) using the 10m Keck telescopes.
The discussion below will be
limited to systems at $z>1.6$ given the dearth of measurements
at lower redshift. A more detailed  review
of DLA abundances in the context of galactic chemical evolution
is given by Lu, Sargent, \& Barlow (1998).

 As can be seen from Figure 1, typical DLA galaxies at $z>1.6$
have $-2.5<$[Fe/H]$<-1$, corresponding to 1/300 to 1/10 solar 
metallicity.  The N(\textsc{H~i})-weighted mean metallicity
is $<$[Fe/H]$>\simeq-1.5$ at $\langle z\rangle=2.5$. The
low metallicities are consistent with them being young galaxies in the
early stages of chemical enrichment. If DLA systems eventually
evolve to solar mean metallicity at $z=0$, their low metallicities
at $z>2$ suggest that most of the baryonic matter in these galaxies
should be in the gas phase rather than in stars and that most of the
star formation should occur at $z<2$, consistent with results from
deep galaxy redshift surveys (cf, Connolly et al 1997).
The N(H~\textsc{i})-weighted mean metallicity in terms of Zn
is about a factor of 2-3 higher than that of Fe (Pettini et al 1997b),
possibly suggesting that some of the Fe atoms are locked up in
dust grains in the DLA galaxies.

 The mean metallicity of DLA systems clearly increases
from $z>4$ to $z\sim 2$ as expected.
However, there is a factor of $\sim$30 scatter
in [Fe/H] at $2<z<3$, presumably reflecting differences in the
formation epoch/star formation history of the galaxies and/or a mixture
of morphological types.  The metallicity distribution
appears to reach  a ``plateau'' value of [Fe/H]$\sim -2$ to $-2.5$ at $z>4$.
Coincidentally, this ``plateau'' metallicity
is identical (within the measurement uncertainties) to that  found for the
intergalactic medium  clouds at similar redshifts, as inferred
from the \textsc{C~iv} absorption associated with Ly$\alpha$ forest
clouds (Cowie et al 1995; Tytler et al 1995;
Songaila \& Cowie 1996).  This coincidence
suggests that the metals in DLA galaxies with [Fe/H]$\sim -2$ to $-2.5$
may simply reflect those in the intergalactic medium,
 possibly produced by Pop III stars
(Ostriker \& Snedin 1996).
If this interpretation is correct, then significant star formation did
not start in DLA galaxies until $z\sim 3-4$.
Such an inference is consistent with the decline in the neutral gas
content of DLA systems at $z>3$ (Storrie-Lombardi, McMahon, \& Irwin 1996)
(presumably because DLA galaxies
are still being formed at such high redshifts) and with the rapid decline
in the space density of quasars at $z>3$ (Schmidt, Schneider, \& Gunn 1995).
It will be important to study more DLA systems at the highest redshift
possible
to confirm the reality of the ``plateau'' metallicity, and to improve
the accuracy of the metallicity determination for the
Ly$\alpha$ forest clouds, which at present may be uncertain 
by as much as a factor of 10 due to uncertain ionization corrections.
 
\section{Dust in Damped Ly$\alpha$ Systems}

   The low heavy element abundances in DLA systems provide a
fundamental limitation to the amount of dust that can form 
in these galaxies: the highest dust-to-gas ratio that a galaxy with 
metallicity $Z$ can have is $Z/Z_{\odot}$ times the Galactic
dust-to-gas ratio,  when  all the atoms from one or more elements
have been incorporated into grains\footnote{Here we ignore
complications resulting from non-solar relative
abundances or from different compositions of dust grains.}.
Since DLA systems at $z>1.6$ generally have [Fe/H]=$-2.5$ to $-1$
or [Zn/H]=$-2$ to $-0.5$ (section 2), it can be concluded without
any detailed analysis that typical DLA systems should
have dust-to-gas ratios in the range of 0.01 to 0.1 times the
Galactic value,  or lower.

    The presence of dust in DLA systems has been inferred from 
the modest reddening of quasars with DLA absorption
in their spectra compared to those without 
(Pei, Fall, \& Bechtold 1990), and from the
sub-solar Cr/Zn (and Fe/Zn) ratios found in DLA systems 
which can be interpreted
as a consequence of dust depletion (Meyer \& Roth 1990;
Pettini et al 1994). However, the extent to which DLA abundances
are significantly affected by dust depletion and the nature of dust
are still controversial (see, for example, Lu et al
1997 for a discussion and for references). If it is assumed that
dust grains in DLA systems have the same composition as Galactic
dust, then a typical dust-to-metal ratio of 40-80\% of the Galactic
value is deduced (Kulkarni, Fall, \& Truran 1997; Pettini et al 1997a; 
Vladilo 1997). Consequently, typical DLA systems should have dust-to-gas 
ratios in the range 0.01-0.1 times the Galactic value.

    DLA systems are defined to have N(\textsc{H~i})$\ge 10^{20}$ cm$^{-2}$
and are observed to have N(\textsc{H~i}) upto $5\times 10^{21}$ cm$^{-2}$.
Adopting the Galactic extinction curve with
E$_{B-V}=$N(\textsc{H~i})/$5\times 10^{21}$, the expected
reddening from DLA systems should be E$_{B-V}\sim0.1$
for N(\textsc{H~i})=$5\times 10^{21}$ cm$^{-2}$ and 
$Z\sim0.1Z_{\odot}$, and should be much less  for lower 
N(\textsc{H~i}) and/or metallicity systems. Adopting Magellanic
Clouds-type extinction curves, 
E$_{B-V}=$N(\textsc{H~i})/$2\times 10^{22}$ for LMC and
E$_{B-V}=$N(\textsc{H~i})/$10^{23}$ for SMC (Koornneef 1984),
would result in even less reddening.

\section{CO Molecules in Damped Ly$\alpha$ Systems}

  Molecules can have important effects on the chemical and physical properties
of interstellar gas. In the Galactic interstellar medium,
molecular hydrogen (H$_2$) and carbon monoxide (CO) have 
been particularly well studied through their UV absorption lines
(Liszt 1997).  Molecular hydrogen has numerous UV transitions 
at $\lambda<1100$ \AA\ from the Lyman and Werner bands. Similarly, CO 
has many UV
transitions at $\lambda<1550$ \AA.  At $z>1.6$, many of these transitions
are shifted into the optical wavelength, enabling their study from the ground.

Absorption from H$_2$ has been detected in only a few DLA systems. In general,
the fraction of H$_2$ relative to total H is small compared to
interstellar clouds in the Milky Way.  This topic is discussed in detail 
by Ge (1997).

Wiklind \& Combes (1994) searched for CO emission at millimeter
wavelength from 8 DLA systems (7 at $z>1.9$) and failed to detect any.
They also reported non-detections of CO absorption 
(N(CO)$<4\times 10^{15}$ cm$^{-2}$) from two of these
systems. In general, CO emission  traces dense molecular clouds.
Absorption lines of CO in the UV provide a much more sensitive
means to probe diffuse interstellar gas, reaching N(CO) 
as low as $10^{12}$ cm$^{-2}$.  A few nondetections of CO in 
DLA systems have been reported previously
(Black, Chaffee, \& Foltz 1987; Chaffee, Black, \& Foltz 1988;
Levshakov et al 1992; Ge et al 1997).
We have searched for the 6 strongest
CO $A^1\Pi-X^1\Sigma^+$ transitions (0-0,1-0,2-0,3-0,4-0,5-0, 6-0) at
  rest-frame wavelength 1544.451 ($f=1.56\times10^{-2}$),
                   1509.750 ($f=3.43\times10^{-2}$),
                   1477.568 ($f=4.12\times10^{-2}$),
                   1447.355 ($f=3.61\times10^{-2}$),
                   1419.046 ($f=2.58\times10^{-2}$),
              and  1392.525 \AA ($f=1.61\times10^{-2}$)
in 17 DLA systems for which we have Keck High Resolution Spectrometer
(HIRES) observations.
No CO absorption is convincingly detected in any of the systems,
resulting in the 2$\sigma$ upper limits given in Table 1. 
These CO limits are roughly comparable to the values deduced for
diffuse interstellar clouds in the Milky Way (eg, Federman et al 1980).

\begin{figure}
\centerline{\vbox{ \psfig{figure=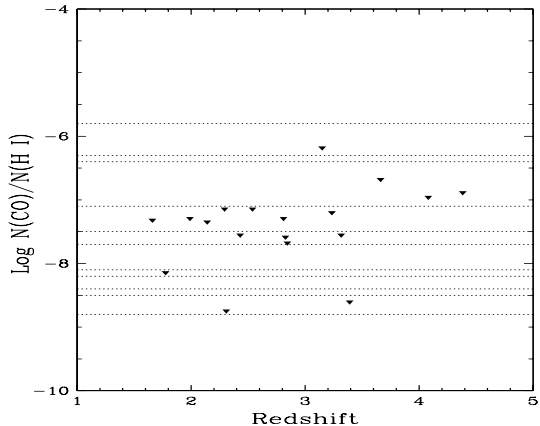,height=2.5in,width=3.0in} }}
\caption[]{CO/\textsc{H~i} ratio vs redshift for damped Ly$\alpha$ systems
(solid triangles). All data points displayed are 2$\sigma$ upper limits.
The horizontal dotted lines indicate the CO/H ratios in the sample
of Galactic sightlines studied by Federman et al (1980), where the highest
ratio is for the $\zeta$ Oph cloud.}
\end{figure}

\begin{table}[htb]
\begin{center}
\caption{Abundances of CO Molecules in Damped Ly$\alpha$ Systems}
\begin{tabular}{lccccr}
\hline
\ \ \ QSO &$z_{DLA}$ &log N(\textsc{H i}) &log N(CO)$^a$ 
&N(CO)/N(\textsc{H~i})$^b$ &N(\textsc{C~ii}$^*$)$^c$ \\
\hline
0000$-$2620 &3.3901    &21.41  &$\leq12.80$  &$\leq2.5\times 10^{-9}$ 
   &...   \\
0100$+$1300 &2.3090    &21.32  &$\leq12.58$  &$\leq1.8\times 10^{-9}$ 
   &13.59 \\
0216$+$0803 &2.2931    &20.45  &$\leq13.31$  &$\leq7.2\times 10^{-8}$ 
   &...   \\
0528$-$2505 &2.8110    &21.20  &$\leq13.91$  &$\leq5.1\times 10^{-8}$ 
   &14.22 \\
0528$-$2505 &2.1410    &20.70  &$\leq13.35$  &$\leq4.5\times 10^{-8}$ 
   &... \\
0930$+$2858 &3.2352    &20.18  &$\leq12.98$  &$\leq6.3\times 10^{-8}$ 
   &$\leq12.53$ \\
1055$+$4611 &3.3172    &20.34  &$\leq12.78$  &$\leq2.8\times 10^{-8}$ 
   &... \\
1104$-$1804A &1.6614    &20.80  &$\leq13.48$  &$\leq4.8\times 10^{-8}$
   &... \\
1202$-$0725 &4.3829    &20.60  &$\leq13.71$  &$\leq1.3\times 10^{-7}$ 
   &$\leq13.06$ \\
1331$+$1704 &1.7764    &21.18  &$\leq13.04$  &$\leq7.2\times 10^{-9}$ 
   &... \\
1425$+$6039 &2.8268    &20.30  &$\leq12.72$  &$\leq2.6\times 10^{-8}$ 
   &... \\
1850$+$4015 &1.9898    &21.60  &$\leq14.31$  &$\leq5.1\times 10^{-8}$ 
   &... \\
1946$+$7658 &2.8443    &20.27  &$\leq12.59$  &$\leq2.1\times 10^{-8}$ 
   &$\leq12.46$ \\
2212$-$1626 &3.6617    &20.20  &$\leq13.52$  &$\leq2.1\times 10^{-7}$ 
   &$\leq12.90$ \\
2233$+$1310 &3.1493    &20.00  &$\leq13.82$  &$\leq6.6\times 10^{-7}$ 
   &$\leq13.53$ \\
2237$-$0608 &4.0803    &20.52  &$\leq13.56$  &$\leq1.1\times 10^{-7}$ 
   &$\leq12.53$ \\
2343$+$1232 &2.4313    &20.34  &$\leq12.78$  &$\leq2.8\times 10^{-8}$ 
   &12.77 \\
2344$+$1228 &2.5379    &20.36  &$\leq13.22$  &$\leq7.2\times 10^{-8}$ 
   &$\leq12.95$ \\
\hline
\end{tabular}
\end{center}
$^a$ All CO limits are 2$\sigma$ estimates based on our Keck 
observations, except for the Q1331+1704 DLA system which is from 
Chaffee et al (1988). 

$^b$ Abundance of CO relative to \textsc{H~i}, which may be taken as the 
abundance of CO relative to the total hydrogen since
DLA systems contain negligible amount of H$_2$ (Ge 1997).
 
$^c$ Based on either our unpublished Keck observations or Table 18 of Lu
et al (1996). All upper limits are 2$\sigma$.

\end{table}

In Figure 2 the CO abundances in DLA systems are compared with those found in
diffuse interstellar clouds in the Milky Way galaxy (Federman et al 1980),
where we take 
N(CO)/N(H)$\equiv$N(CO)/(N(\textsc{H~i})+2N(H$_2$))$\simeq$N(CO)/N(\textsc{H~i})
since the abundance of H$_2$ is generally negligible in DLA systems 
(Ge 1997).
The CO/H upper limits are broadly consistent with observations
of Galactic diffuse interstellar clouds, especially considering the
low metallicities of DLA systems which should reduce the CO abundances
in gas with the same physical conditions.  The highest CO/H ratio found
in Galactic interstellar clouds is toward $\zeta$ Oph, with 
CO/H$\simeq 1.6\times 10^{-6}$ (Ironically, the $\zeta$ Oph cloud
is often quoted as a representative diffuse interstellar cloud).

In the Galactic interstellar medium,
significant CO absorption (N(CO)$>10^{12}$ cm$^{-2}$)
is detected only in sightlines with N(H$_2$)$>10^{19}$
cm$^{-2}$ (Federman et al 1980).  The main
destruction mechanism of H$_2$ and CO in diffuse clouds is  photodissociation
by UV radiation. Once the column density of H$_2$ reaches some 
threshold value, self-shielding in the
Lyman and Werner bands will prevent UV photons
from penetrating the interior of the clouds, and the abundances
of H$_2$ and CO build up quickly. This threshold H$_2$ column density
appears to be around N(H$_2$)$\sim 10^{19}$ cm$^{-2}$ in the
Galactic ISM, above which the abundance of H$_2$, N(H$_2$)/N(H),
jumps abruptly from $10^{-6}-10^{-2}$ to $\sim10^{-1}$ (Savage et al 1977). 
Sightlines with such high N(H$_2$) generally have N(H)$>5\times 10^{20}$ 
cm$^{-2}$ and are fairly reddened (E$_{B-V}>0.1$).
DLA systems, on the other hand, have very low H$_2$ abundances
(N(H$_2$)$<10^{19}$ cm$^{-2}$; Ge 1997),
 which is probably a direct consequence of their 
low dust-to-gas ratio (section 3).  In these regards,
the non-detection of CO in DLA systems is probably not too
surprising. The low CO abundances in DLA systems may also be
affected by the low metallicities of these systems.

\section{[C II] Emssion from Damped Ly$\alpha$ Systems}
  
   The [\textsc{C~ii}] 158$\mu$m far-infrared transition, which
results from the two fine-structure levels of the ground state
of the \textsc{C~ii} ion, is an important cooling agent for
gas in the cool neutral and warm ionized medium (Kulkarni \& Heiles 1987). 
A significant fraction of the far-infrared luminosity of nearby
galaxies is contained in this line emission,
hence the strength of the [\textsc{C~ii}]
line can be a potentially important diagnostic of the physical
conditions in high redshift galaxies (eg, Phillips 1997; van den Werf 1997). 

The cooling rate resulting
from the [\textsc{C~ii}] line can be estimated from the column
density of ions in the excited fine structure level, 
N(\textsc{C~ii}$^*$), through the relation (cf, Savage et al 1993)
$$l_c=h\nu_{12}{\rm N(C II}^*)A_{21}/{\rm N(H I)},$$
where $A_{21}=2.36\times 10^{-6}$ s$^{-1}$ is the Einstein $A$-coefficient.
The column density of N(\textsc{C~ii}$^*$)
can be estimated from the \textsc{C~ii}$^*$ $\lambda$1335.708 UV 
transition. The last column of Table 1 gives the N(\textsc{C~ii}$^*$)
estimates or 2$\sigma$ upper limits in DLA systems based on our Keck HIRES
observations. The resulting cooling rates are 
$l_c$=$(6, 30, 8)\times 10^{-28}$ ergs s$^{-1}$ 
hydrogen$^{-1}$ for the three measurements, and the
corresponding  upper limits are in the range of 
$3-100 \times 10^{-28}$ ergs s$^{-1}$ hydrogen$^{-1}$.
These measurements are at least two orders of magnitude lower
than the average value found by Pottasch, Wesselius, \& van 
Duinen (1979) for eight different sightlines through the Galactic
disk. If we assume that a typical DLA galaxy contains 
$10^{10}M_{10}$ solar mass of hydrogen and that the
DLA sightline provides a fair sample of the entire galaxy, the total
luminosity contained in the [\textsc{C~ii}] line would be
only $10^6-10^7 M_{10} L_{\odot}$ based on the three measurements.
These values are several orders of magnitude lower 
than the detection limit achievable
with current instruments (cf, van den Werf 1997).

\section{Words of Caution}

   The properties of DLA systems discussed above are based on the observed
population of DLA systems. There are a number of factors that could 
potentially bias our view of these high redshift galaxies, including:

(1) More dusty (presumably more metal-rich also) galaxies should have
a stronger effect dimming the background quasars through dust
obscuration. Hence surveys of
DLA systems using magnitude-limited quasar samples may preferentially
exclude such dusty galaxies from the sample. The effect of such
a selection bias is probably relatively small at $z>1.6$ but could
be large at lower redshift (Fall 1997). A direct check of this issue
can be made by carrying out a survey of DLA systems using a complete
sample of radio-selected quasars without regard to their optical
brightness.
 
(2) The sampling rate provided by DLA systems 
of different morphological type of galaxies
or different regions of the same galaxy 
depends on the relative absorption cross sections of these galaxies
or regions.  For example, the outer regions of disk galaxies
will be much more heavily represented by DLA systems than the inner
regions of disks. Hence properties revealed by DLA systems may not
be representative of the overall extent of galaxies.

\acknowledgements{We thank
Art Wolfe and Jason X. Prochaska for providing
their abundance measurements in advance of publication.}


\begin{references}

\reference
Black, J.H., Chaffee, F.H. Jr., \& Foltz, C.B. 1987, ApJ, 317, 442

\reference
Briggs, F.H. 1988, in QSO Absorption Lines: Probing the Universe,
 eds. J.C. Blades, D.A. Turnshek, \& C.A. Norman (Cambridge University
 Press), p275
 
\reference
Chaffee, F.H.Jr., Black, J.H., \& Fotz, C.B. 1988, ApJ, 335, 584

\reference
Connolly, A.J., Szalay, A.S., Dickinson, M., SubbaRao, M.U., Brunner, R.J.
      1997, ApJ, 486, L11
 
\reference
 Cowie, L.L., Songaila, A., Kim, T.S., \& Hu, E.M. 1995, AJ, 109, 1522
 
\reference
Fall, S.M. 1997, this volume

\reference
Federman, S.R., Glassgold, A.E., Jenkins, E.B., \& Shaya, E.J. 1980,
   ApJ, 242, 545

\reference
 Ge, J. 1997, this volume
 
\reference
Ge, J., Bechtold, J., Walker, C., \& Black, J.H. 1997, ApJ, 486, 727
 
\reference
Koornneef, J. 1984, in Structure and Evolution of the Magellanic Clouds,
  eds. S. van den Berg \& K.S. de Boer (Kluwer), p133

\reference
Kulkarni, S.R., \& Heiles, C. 1987, in Interstellar Processes,
  eds. D.J. Hollenbach \& H.A. Thronson, Jr. (Dordrecht:Reidel), p87

\reference
Kulkarni, V.P., Fall, S.M., \& Truran, J.W. 1997, ApJ, 484, L7
 
\reference
Lanzetta, K.M., Wolfe, A.M., \& Turnshek, D.A. 1995, ApJS, 440, 435
 
\reference
Levshakov, S.A., Chaffee, F.H. Jr., Foltz, C.B., \& Black, J.H. 1992, 
   A\&A, 262, 385

\reference
Liszt, H. 1997, this volume

\reference
Lu, L., Sargent, W.L.W., \& Barlow, T.A. 1998, in Cosmic Chemical
   Evolution, IAU Symp. 187, in press
 
\reference
Lu, L., Sargent, W.L.W, Barlow, T.A., Churchill, C.W., \&
    Vogt, S. 1996, ApJS, 107, 475
 
\reference
Meyer, D.M., \& Roth, K.C. 1990, ApJ, 363, 57
 
\reference
Ostriker, J.P., \& Snedin, N.Y. 1996, ApJ, 472, L63
 
\reference
Pei, Y.C., Fall, S.M., \& Bechtold, J. 1991, ApJ, 378, 6
 
\reference
 Pettini, M., Smith, L.J., Hunstead, R.W., \& King, D.L.
    1994, ApJ, 426, 79
 
\reference
 Pettini, M., King, D.L., Smith, L.J.,\& Hunstead, R.W.
    1997, ApJ, 478, 536 (a)
 
\reference
 Pettini, M., Smith, L.J., King, D.L., \& Hunstead, R.W.
    1997, ApJ, 486, 665 (b)
 
\reference
Phillips, T. 1997, this volume

\reference
Pottasch, S.R., Wesselius, P.R.,  \& van Duinen, R.J. 1979
  A\&A, 74, L15

\reference
Savage, B.D., Bohlin, R.C., Drake, J.F., \& Budich, W. 1977,
   ApJ, 216, 291

\reference
Savage, B.D., Lu, L., Weymann, R.J., Morris, S.L., \& Billiland, R.L.
  1993, ApJ, 404, 124

\reference
 Schmidt, M., Schneider, D.P., \& Gunn, J.E. 1995, AJ, 100, 68
 
\reference
Smith, H.E., Jura, M., \& Margon, B. 1979, ApJ, 228, 369

\reference
Songaila, A, \& Cowie, L.L. 1996, AJ, 112, 335
 
\reference
 Storrie-Lombardi, L.J., McMahon, R.G., \& Irwin, M.J. 1996
     MNRAS, 283, L79
 
\reference
 Tytler, D., Fan, X.-M., Burles, S., Cottrell, L., Davis, C.,
   Kirkman, D., \& Zuo, L. 1995, in {\it QSO Absorption Lines}, ed. G.Meylan
   (Springer-Verlag), p289
 
\reference
van den Werf, P. 1997, this volume

\reference
Vladilo, G. 1997, ApJ, in press
 
\reference
Wiklind, T., \& Combes, F. 1994, A\&A, 288, L41

\reference
 Wolfe, A.M. 1988, in QSO Absorption Lines: Probing the
  Universe, eds.  Blades, Turnshek, and Norman
  (Cambridge Univ Press), p297
 
\end{references}
\end{document}